\documentclass[multicol,prb,twocolumn,showpacs,aps]{revtex4}
\usepackage{amsmath}
\usepackage{array}
\usepackage{graphicx}
\usepackage{color}
\usepackage{subfigure}

\begin{document}

\title{Coexistence of p-wave Cooper pairing and ferromagnetism}

\author{Xu Yan}

\author{Qiang Gu}
\email{qgu@ustb.edu.cn}
\affiliation{Department of Physics
University of Science and Technology Beijing, Beijing 100083, China}

\date{\today}

\begin{abstract}

A two-band model for coexistence of p-wave superconductivity with
localized ferromagnetism is studied using the equation of motion
approach. It shows that ferromagnetic and superconducting states
enhance each other but in a different way from that of the one-band
model. The Curie temperature is not only determined by the exchange
interactions between localized spins, but also can be increased with
the coupling between electrons and spins, and with the p-wave
Cooper-pairing interaction. These results are complementary to those
of the one-band model, which suggest that the Curie temperature is
unlikely to ever be below the superconducting transition
temperature.



\end{abstract}

\maketitle

\section{Introduction}\label{intro}

Motivated by the recent discoveries of the ferromagnetic
superconductors, e.g., ${\rm UGe_{2}}$ \cite{saxena2000} and URhGe
\cite{aoki2001}, much attention has been drawn to understanding the
underlying physics of the coexistence of superconductivity and
ferromagnetism
\cite{karchev2001,suhl2001,Walker2002,kirkpatrick2004,nevidomskyy2005,jian2009}.
Early investigations supposed that electrons form conventional
s-wave Cooper pairs \cite{karchev2001,suhl2001}. At present, the
scenario of spin-triplet p-wave pairing is generally accepted
\cite{Walker2002,kirkpatrick2004,nevidomskyy2005,jian2009}.

Nevidomskyy proposed a microscopic model of the coexistence of a
p-wave spin-triplet superconductivity with weakly itinerant
ferromagnetism \cite{nevidomskyy2005}. Supposing that the coexisting
state is a uniform Meissner state, he explained the enhancement of
superconductivity by the established ferromagnetism. On the
other hand, Jian {\it et al.} studied the feedback effect of
superconductivity upon ferromagnetism \cite{jian2009}. Due to the
interplay between the ferromagnetic order and p-wave Cooper pairing,
it was suggested that the Curie temperature is unlikely to ever be
below the superconducting transition temperature once the
ferromagnetism is established. These results are to some extent
consistent with the observed phase diagram of ${\rm UGe_{2}}$
\cite{saxena2000} and with the theoretical discussion of Walker and
Samokhin \cite{Walker2002}.

It is now well-believed that the electrons involved in both the
ferromagnetic (FM) and superconducting (SC) orders are within the
same itinerant band. Nevertheless, it helps for arriving at a
complete understanding of ferromagnetic superconductivity to
investigate the coexistence of superconductivity and localized
magnetic order. Suhl proposed a mechanism of simultaneous appearance
of ferromagnetism and superconductivity based on interactions
between electrons mediated by localized spins \cite{suhl2001}. More
recently, Singh discussed a model consisting of a pairing
interaction and a term describing the scattering of Cooper pairs by
localized electrons \cite{singh2011}.

Here we investigate the interplay between FM and SC orders based on
a two-band model. A band of itinerant electrons which can exhibit
the A-phase p-wave Cooper pairing is coupled to a lattice of
localized spins with FM couplings. The model is treated using
equations of motion truncated at the lowest nontrivial order. It is
shown that ferromagnetism and superconductivity affect on each other
in a different way from that of the one-band model.

\section{The model}\label{model}
We use the Heisenberg model on a simple cubic lattice to describe
localized spins,
\begin{align}
H_S = - J\sum_{\langle i,j\rangle} {\bf S}_i\cdot {\bf S}_j .
\end{align}
The superconducting electrons are described by the BCS Hamiltonian,
\begin{align}
H_e&= \sum_{\bf k\sigma}(\varepsilon_{\bf k}-\mu) a_{\bf k\sigma}^\dagger a_{\bf k\sigma}\nonumber \\
  & + \frac{1}{2v}\sum_{\bf k \bf k',\sigma\sigma'} V_{\bf k,\bf k'}
    a_{\bf k,\sigma}^\dagger a_{\bf -k,\sigma'}^\dagger
    a_{\bf -k'\sigma'}a_{\bf k'\sigma} .
\end{align}
There exists an exchange interaction between localized spins and
itinerant electrons,
\begin{align}
H_{I}&=-g\sum_{j}\{S_{j}^{z}(n_{j\uparrow}-n_{j\downarrow})+S_{j}^+
  a_{j\downarrow}^\dagger a_{j\uparrow}+S_{j}^- a_{j\uparrow}^\dagger
  a_{j\downarrow}\} .
\end{align}
Here ${\bf S}_{i}$ represents the spin operator at site $i$. $J$ is
the exchange integral between localized spins, and $J>0$ for the FM
state. $a_{\bf k\sigma}^\dagger$($a_{\bf k\sigma}$) is the creation
(annihilation) operator of electrons. We discuss the case of
ferromagnetic coupling between electrons and spins, so the coupling
strength $g$ is positive. The pairing potential is assumed to have
the p-wave type, $V_{\bf k,\bf k'}=-V \hat{\bf k}\cdot \hat{\bf
k'}$, and here we choose the SC order parameters to have the
$A$-phase symmetry \cite{nevidomskyy2005,jian2009},
$\Delta_{\pm}({\bf k})=(\hat{k}_x+i \hat{k}_y)\Delta_{\pm}$.

The Hamiltonian is dealt with using the Green's function method
within the mean-field theory framework. The Green's functions are
defined as follows,
\begin{align}
\ll S_{\bf k}^+(t); S_{\bf -k}^- \gg=-i\Theta(t)\langle[S_{\bf k}^+(t),S_{\bf -k}^{-}]\rangle ,\\
\ll a_{\bf k\sigma}(t); a_{\bf k\sigma}^\dagger \gg=-i\Theta(t)\langle[a_{\bf k\sigma}(t), a_{\bf k\sigma}^\dagger]\rangle  ,\\
\ll a_{\bf k\sigma}^\dagger(t); a_{\bf -k\sigma}^\dagger \gg=-i\Theta(t)\langle[a_{\bf k\sigma}^\dagger(t), a_{\bf -k\sigma}^\dagger]\rangle .
\end{align}
Using the equations of motion approach and truncation technique, we
get Green's functions at the lowest nontrivial order and derive the
self-consistent equations,
\begin{align}
M=&\left[2+\frac{1}{2\pi^3}\int_{-\pi}^{\pi}dk_x \int_{-\pi}^{\pi}dk_y \int_{-\pi}^{\pi}dk_z  \right.\nonumber \\
&\left. \left(e^{\frac{ 2\overline{J}M\left(3-\cos{k_x}-\cos{k_y
}-\cos{k_z }\right)+\overline{g} m}{\overline{T}}}-1\right)^{-1}\right]^{-1} ,
\end{align}
\begin{align}
n=&\frac{3}{16}\int_{0}^{\infty} d\overline{\varepsilon} \int_{0}^{\pi}d\theta \sqrt{\overline{\varepsilon}} \sin{\theta} \nonumber \\
&\times \left(2-\frac{\overline{\omega_1}
\tanh\left(\frac{\overline{a}}{2\overline{T}}\right)}{\overline{a}}
-\frac{\overline{\omega_2}
\tanh\left(\frac{\overline{b}}{2\overline{T}}\right)}{\overline{b}}\right) ,
\end{align}
\begin{align}
m=&\frac{3}{16}\int_{0}^{\infty} d\overline{\varepsilon} \int_{0}^{\pi}d\theta \sqrt{\overline{\varepsilon}} \sin{\theta} \nonumber \\
&\times \left(\frac{\overline{\omega_2}
\tanh\left(\frac{\overline{b}}{2\overline{T}}\right)}{\overline{b}}-\frac{\overline{\omega_1}
\tanh\left(\frac{\overline{a}}{2\overline{T}}\right)}{\overline{a}}\right) ,
\end{align}
\begin{align}
1=&\frac{3\overline{V}}{32}\int_{\overline{\mu}+\overline{g}M-\omega_c}
^{\overline{\mu}+\overline{g}M+\omega_c} d\overline{\varepsilon} \int_{0}^{\pi} d\theta \sqrt{\overline{\varepsilon}} \sin^3{\theta}\frac{\tanh\left(\frac{\overline{a}}{2\overline{T}}\right)}{\overline{a}} ,
\end{align}
\begin{align}
1=&\frac{3\overline{V}}{32}
\int_{\overline{\mu}-\overline{g}M-\omega_c}^{\overline{\mu}-\overline{g}M+\omega_c} d\overline{\varepsilon} \int_{0}^{\pi} d\theta \sqrt{\overline{\varepsilon}} \sin^3{\theta}\frac{\tanh\left(\frac{\overline{b}}{2\overline{T}}\right)}{\overline{b}} ,
\end{align}
where we define $M=\langle S^z \rangle $, and
$\overline{a}=\sqrt{\overline{\omega}_1^2+\overline{\Delta}_{+}^2
\sin^2{\theta}}$,
$\overline{b}=\sqrt{\overline{\omega}_2^2+\overline{\Delta}_{-}^2
\sin^2{\theta}}$,
$\overline{\omega}_1=\overline{\varepsilon}-\overline{\mu}-\overline{g}M$,
$\overline{\omega}_2=\overline{\varepsilon}-\overline{\mu}+\overline{g}M$.
Here all the parameters are nondimensionalized, for example,
$\overline{g}=g/\epsilon_f$, $\overline{V}=V/\epsilon_f$. Other
dimensionless parameters, $\overline{\varepsilon}$,
$\overline{\mu}$, $\overline{\omega}$, $\overline{\Delta}_{\pm}$,
and $\overline{\omega}_c$ are rescaled analogously. The rescaled
factor $\epsilon_f=\frac{\hbar^2}{2m} (3\pi^2 n)^{2\over3}$ and
$n=1$ at half-filling. The dimensionless temperature is defined as
$\overline{T}=k_B T/\epsilon_f$. The energy cutoff
$\overline{\omega}_{c}=0.01\overline{\varepsilon}_{F}$ is chosen to
be consistent with the one-band model \cite{jian2009}, where
$\overline{\varepsilon}_{F}$ is the dimensionless Fermi energy.

\begin{figure}
\begin{center}
\includegraphics[width=0.4\textwidth]{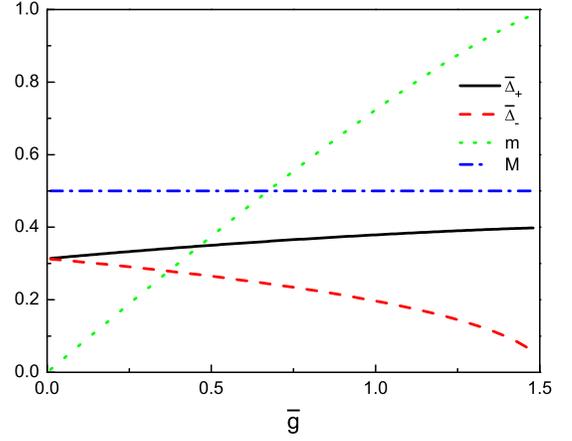}
\end{center}
\caption{Plots of SC gaps $\overline{\Delta}_{\pm}$,
itinerant-electron magnetization $m$, and local-spin magnetization
$M$, as a function of the exchange interaction between electrons and
spins, $\overline{g}$. Here $\overline{V}=100$ and
$\overline{J}=0.1$ } \label{fig1}
\end{figure}

\begin{figure}
\centering
\includegraphics[width=0.4\textwidth]{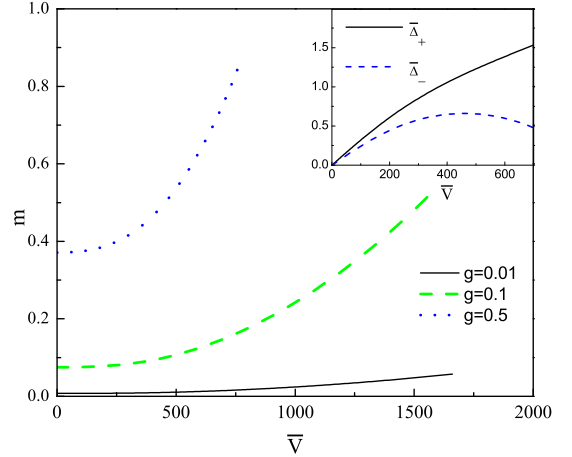}
\caption{Plot of magnetization density of itinerant electrons
$\overline{m}$ as a function of p-wave interaction strength
$\overline{V}$ at $\overline{T}=0$, $\overline{J}=0.01$. Inset: Plot
of superconducting gaps as function of $\overline{V}$ with
$\overline{g}=0.5$.} \label{fig2}
\end{figure}

\begin{figure}[b]
\begin{center}
\includegraphics[width=0.4\textwidth]{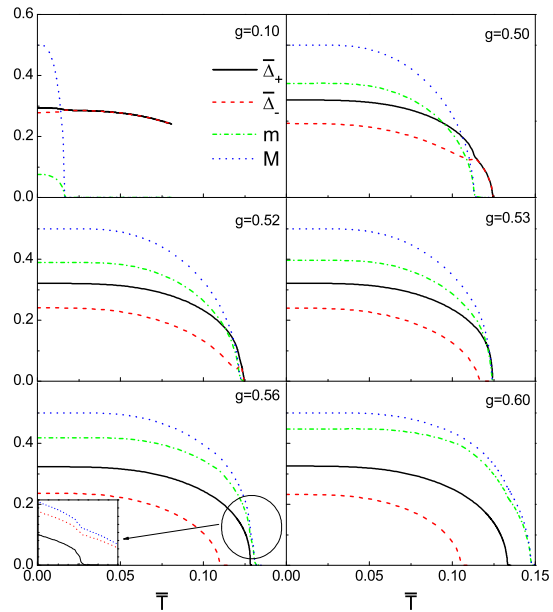}
\end{center}
\caption{Plots of all order parameters, $\overline{\Delta}_{\pm}$,
$m$, and $M$, as functions of $\overline{T}$ with
$\overline{V}=100$, $\overline{J}=0.01$, at different $\overline{g}$
values. Inset: Enlargement of the region $0.125<\overline{T}<0.13$.}
\label{fig3}
\end{figure}

\section{Results and discussion}\label{results}

We first calculate the $T=0$ properties. Figure \ref{fig1} plots the
p-wave SC order parameters, $\overline{\Delta}_{\pm}$, magnetization
density of itinerant electrons, $m=\langle n_{+}\rangle-\langle
n_{-} \rangle$, and magnetization of localized spins, $M=\langle S^z
\rangle $. $m$ is improved as the exchange interaction
$\overline{g}$ increases. Correspondingly, the SC gap
$\overline{\Delta}_{+}$ is strengthened, while
$\overline{\Delta}_{-}$ is weakened.

Figure \ref{fig2} shows the variation of $m$ and
$\overline{\Delta}_{\pm}$ with the p-wave interaction strength
$\overline{V}$. Apparently, $m$ rises as $\overline{V}$ increases
for each given value of $\overline{g}$, indicating that p-wave
Cooper pairing can also enhance the ferromagnetism. As shown in the
inset, with increasing $\overline{V}$, $\overline{\Delta}_{+}$ rises
accordingly, while $\overline{\Delta}_{-}$ initially rises and then
decreases. These results are consistent with the one-band model
\cite{jian2009}.

Figure \ref{fig3} illustrates all the order parameters at finite
temperatures. It seems that the FM transition temperature
($\overline{T}_{FM}$) is mainly determined by the exchange
interaction between localized spins, $J$, when the coupling between
localized spins and itinerant electrons, $\overline{g}$, is small.
So the FM state can vanish earlier than the SC state as long as $J$
is weak enough. Once $m$ and $M$ disappear, $\overline{\Delta}_{+}$,
$\overline{\Delta}_{-}$ become equal. $\overline{T}_{FM}$ rises as
$\overline{g}$ increases and it tends to $\overline{T}_{SC}$ as
$\overline{g}$ goes up to about $0.52$. And then $\overline{T}_{FM}$
surpasses $\overline{T}_{SC}$ at larger $\overline{g}$ values. These results
are different from those of the one-band model for which
$\overline{T}_{FM}$ is unlikely below $\overline{T}_{SC}$ once the
ferromagnetism is established \cite{jian2009}. Figure \ref{fig3}
also indicates that the magnetization displays an inflexion at the
SC transition temperature.

\section{Conclusion}\label{concl}

In conclusion, we study a two-band model describing coexistence of
p-wave superconductivity with localized ferromagnetism. It is shown
that ferromagnetism can be enhanced by the p-wave Cooper pairing, as
suggested in a one-band model previously. However, the Curie
temperature in this model is not only determined by the exchange
interaction between localized spins, also by the coupling between
localized spins and itinerant electrons. The Curie temperature can
be lower than the superconducting transition temperature.

\section*{Acknowledgements}
This work is supported by the National Natural Science
Foundation of China (No. 11274039), the Specialized Research Fund for the
Doctoral Program of Higher Education (No. 20100006110021), and the
Fundamental Research Funds for the Central Universities of China.

\end{document}